\documentclass[aoas,rotating,nameyear,dvips]{arximspdf}
\usepackage{dcolumn}
\usepackage{graphicx}

\doi{10.1214/10-AOAS390}
\volume{5}
\issue{1}
\pubyear{2011}
\firstpage{449}
\lastpage{467}

\makeatletter

\let\sv@tabnotetext\tabnotetext
  \let\sv@tabnotemark@fmt\tabnotemark@fmt
   \long\def\legend#1{{\let\tabnote@indent\leavevmode\sv@tabnotetext[]{}{#1}}}

\newcolumntype{d}[1]{D{.}{.}{#1}}

\newcommand{\bfbeta}{\bolds\beta}

\newcommand{\bfSigma}{\bolds\Sigma}
\newcommand{\Var}{\operatorname{Var}}
\newcommand{\age}{\operatorname{age}}
\newcommand{\Corr}{\operatorname{Corr}}
\newcommand{\pr}{\operatorname{pr}}
\newcommand{\smoke}{\operatorname{smoke}}
\newcommand{\HIV}{\operatorname{HIV}}
\newcommand{\wt}{\operatorname{wt}}

\makeatother

\begin{document}
\begin{frontmatter}

\title{A generalized linear mixed model for longitudinal binary data
with a marginal logit link function\thanksref{T1}}

\runtitle{GLMM with marginal logit link}%

\thankstext{T1}{Supported by NIH Grants GM 29745, HL 69800, MH 54693, AI 60373,
CA 74015, CA 70101, CA 69222 and CA 68484.}
\begin{aug}
\author[A]{\fnms{Michael} \snm{Parzen}},
\author[B]{\fnms{Souparno} \snm{Ghosh}},
\author[C]{\fnms{Stuart} \snm{Lipsitz}\corref{}\ead[label=e3]{slipsitz@partners.org}},
\author[D]{\fnms{Debajyoti}~\snm{Sinha}},
\author[E]{\fnms{Garrett M.} \snm{Fitzmaurice}},
\author[F]{\fnms{Bani~K.}~\snm{Mallick}}
\and
\author[G]{\fnms{Joseph G.} \snm{Ibrahim}}

\runauthor{M. Parzen et al.}

\affiliation{Emory University,
Texas A\&M University,
Brigham and Women's Hospital,
Florida State University,
Harvard Medical School,
Texas A\&M University
and
The University of North Carolina at Chapel Hill}

\address[A]{M. Parzen\\
Goizueta Business School\\
Emory University\\
201 Dowman Drive\\
Atlanta, Georgia\\USA} %adresu isvedimo komanda gale!

\address[B]{S. Ghosh\\
Department of Statistics\\
Texas A\&M University\\
College Station, Texas\\
USA}

\address[C]{S. Lipsitz\\
Brigham and Women's Hospital\\
Boston, Massachusetts\\
USA\\
\printead{e3}}

\address[D]{D. Sinha\\
Department of Statistics\\
Florida State University\\
600W. College Avenue\\
Tallahassee, Florida\\
USA}

\address[E]{G. Fitzmaurice\\
Harvard Medical School\\
Boston, Massachusetts\\
USA}

\address[F]{B. K. Mallick\\
Department of Statistics\\
Texas A\&M University\\
College Station, Texas\\
USA}

\address[G]{J. G. Ibrahim\\
Gillings School of Global Public Health\\
The University of North Carolina\\
at Chapel Hill\\
Chapel Hill, North Carolina\\
USA}
\end{aug}

% HISTORY:
\received{\smonth{11} \syear{2009}}
\revised{\smonth{7} \syear{2010}}

% ABSTRACT
%
\begin{abstract}
Longitudinal studies of a binary outcome are common in the health,
social, and
behavioral sciences. In general, a feature of random effects logistic regression
models for longitudinal binary data is that the marginal functional
form, when
integrated over the distribution of the random effects, is no longer of logistic
form. Recently, Wang and Louis [\textit{Biometrika} \textbf{90} (2003)
765--775] proposed a random intercept model in the
clustered binary data setting where the marginal model has a logistic
form. An
acknowledged limitation of their model is that it allows only a single random
effect that varies from cluster to cluster. In this paper we propose a
modification of their model to handle longitudinal data, allowing
separate, but
correlated, random intercepts at each measurement occasion. The
proposed model
allows for a flexible correlation structure among the random
intercepts, where
the correlations can be interpreted in terms of Kendall's $\tau$. For example,
the marginal correlations among the repeated binary outcomes can
decline with
increasing time separation, while the model retains the property of having
matching conditional and marginal logit link functions. Finally, the proposed
method is used to analyze data from a longitudinal study designed to monitor
cardiac abnormalities in children born to HIV-infected women.
\end{abstract}

% KEYWORDS
%

\end{frontmatter}

%s1 ###
\section{Introduction}\label{intro}

Longitudinal studies of a binary outcome are common in the health,
social, and
behavioral sciences. For example, in the Pediatric Pulmonary and Cardiac
Complications (P$^2$C$^2$) of Vertically Transmitted~HIV Infection Study
[Lipshultz et al. (\citeyear{Lipshultz1998})], a longitudinal study
designed to monitor heart disease
and the progression of cardiac abnormalities in children born to HIV-infected
women, a key outcome was the binary variable ``pumping ability of the
heart''
(normal/abnormal). Previous results [Lipshultz et al. (\citeyear
{Lipshultz1998,Lipshultz2000,Lipshultz2002})] from the P$^2$C$^2$ study
have shown that
sub-clinical cardiac abnormalities develop early in children born to
HIV-infected women, and that they are frequent, persistent, and often
progressive. In the P$^2$C$^2$ study a birth cohort of 401 infants born to
women infected with HIV-1 were followed over time for up to six years.
Of these
401 infants, 74 (18.8\%) were HIV positive, and 319 (81.2\%) were HIV
negative. It is of interest to model the effect of HIV status of the
child on
the marginal probability of abnormal pumping ability of the heart over time.
Additional covariates include mother's smoking status during pregnancy,
gestational age and birth-weight standardized for age
(1${}={}$abnormal, 0${}={}$normal).
Table~\ref{tab1} shows data from 10 of the 401 subjects on file.

%t1 ###
%
\begin{table}
\caption{Data from P$^2$C$^2$ longitudinal study for 10 randomly selected
children}\label{tab1}
\tabcolsep=0pt
\begin{tabular*}{\textwidth}{@{\extracolsep{4in minus 4in}}lccccccccccc@{}}
\hline
& & & & &
\multicolumn{7}{c@{}}{\textbf{Heart pumping ability at
age}\tabnoteref[a]{tb1a}\textbf{:}} \\[-5pt]
& & \textbf{Mom} & \textbf{Gest.} & \textbf{Birth}&\multicolumn
{7}{c@{}}{\hrulefill}\\
\textbf{Case}& \textbf{HIV}\tabnoteref[b]{tb1b}&\textbf
{smoked}\tabnoteref[c]{tb1c}
& \textbf{age (weeks)}& \textbf{weight}\tabnoteref[d]{tb1d}& \textbf
{Birth} & \textbf{1} & \textbf{2} & \textbf{3} & \textbf{4}
& \textbf{5} & \textbf{6}
\\
\hline
\hphantom{1}1 & 1 & 0 & 41 & 0 & 0 & 0 & 0 & 0 & 0 & 0 & -- \\
\hphantom{1}2 & 1 & 1 & 34 & 0 & 1 & -- & 0 & 0 & 1 & -- & -- \\
\hphantom{1}3 & 0 & 1 & 40 & 0 & 1 & 0 & 0 & -- & -- & -- & -- \\
\hphantom{1}4 & 1 & 0 & 40 & 0 & 0 & -- & 0 & 0 & 0 & 1 & -- \\
\hphantom{1}5 & 0 & 1 & 39 & 0 & -- & 1 & 0 & -- & -- & -- & -- \\
\hphantom{1}6 & 0 & 1 & 35 & 0 & 1 & -- & -- & -- & -- & -- & -- \\
\hphantom{1}7 & 0 & 0 & 36 & 0 & -- & 0 & 0 & -- & -- & -- & -- \\
\hphantom{1}7 & 1 & 0 & 33 & 1 & -- & 1 & 1 & 1 & -- & -- & -- \\
\hphantom{1}8 & 0 & 0 & 36 & 1 & 0 & 0 & -- & -- & -- & -- & -- \\
\hphantom{1}9 & 0 & 0 & 41 & 1 & -- & -- & -- & -- & 0 & -- & -- \\
10 & 0 & 1 & 34 & 1 & -- & 0 & 0 & -- & 0 & 1 & 0 \\
\hline
\end{tabular*}
\legend{Note: -- $=$ missing.}
\tabnotetext[a]{tb1a}{1 $=$ abnormal, 0 $=$ normal.}
\tabnotetext[b]{tb1b}{1 $=$ HIV positive, 0 $=$ not HIV positive.}
\tabnotetext[c]{tb1c}{1 $=$ mother smoked during pregnancy, 0 mother did not smoke.}
\tabnotetext[d]{tb1d}{1 $=$ low birth-weight for age, 0 $=$ normal birth-weight.}
\end{table}

We consider likelihood-based estimation of the logistic regression
model for the
marginal probabilities of the repeated binary responses. This, of course,
requires a fully parametric likelihood approach based on the joint
multinomial distribution of the repeated binary outcomes from each subject.
In practice, full likelihood-based methods for fitting of marginal
models for
discrete longitudinal data have proven to be very challenging for the following
reasons: (i) it can be conceptually difficult to model higher-order associations
in a flexible and interpretable manner that is consistent with the
model for the
marginal expectations [e.g., Bahadur (\citeyear{Bahadur1961})], (ii)~given a marginal model for the
vector of repeated outcomes, the multinomial probabilities cannot, in general,
be expressed in closed-form as a function of the model parameters, and
(iii) the
number of multinomial probabilities grows exponentially with the number of
repeated measures.

Although various likelihood approaches have been proposed, for example, models
based on two- and higher-order correlations [Bahadur (\citeyear
{Bahadur1961}); Zhao and Prentice
(\citeyear{Zhao1990})] and models based on two- and higher-order odds
ratios [McCullagh and Nelder (\citeyear{McCullagh1989});
Lipsitz, Laird and Harrington (\citeyear{Lipsitz1991}); Molenberghs
and Lesaffre
(\citeyear{Molenberghs1994})], none of these
likelihood-based models have proven to be of real practical use unless the
number of repeated measures is relatively small (say, less than~5). As the
number of repeated measures increases, the number of parameters that
need to be
specified and estimated proliferates rapidly for any of these joint
distributions, and a solution to the likelihood equations quickly becomes
intractable.

Other full likelihood approaches have been formulated as generalized linear
mixed models (GLMM). For example, Heagerty (\citeyear{Heagerty1999})
and Heagerty and Zeger
(\citeyear{Heagerty2000}) have developed a likelihood-based approach
that combines the versatility
of GLMMs for modeling the within-subject association with a marginal logistic
regression model for the marginal probability of response. They refer
to their
general class of models as \textit{marginalized} random effects
models. Recall
that in the standard GLMM for binary outcomes, the marginal probabilities,
obtained by integrating over the random effects, in general, no longer
follow a
generalized linear model, due to the nonlinearity of the link function
typically adopted in regression models for discrete responses. In
contrast, the
\textit{marginalized} random effects model can be specifically
formulated such that
the marginal probabilities follow a logistic regression model. Unlike the
standard generalized linear mixed model, the marginalized random
effects models
of Heagerty (\citeyear{Heagerty1999}) has no closed form expression for
the conditional probability
of response (conditional on the random effects). When the main interest
is in
the marginal model parameters, the latter feature has no impact on the
interpretability of the model; however, it can be a drawback when
trying to
implement an algorithm to obtain the maximum likelihood (ML) estimates using
commonly available software, for example, PROC NLMIXED in SAS (V9.2).

In this paper the goal of our approach is to develop a generalized
linear mixed model which has a straightforward interpretation of the
effect of
the
covariates, both conditionally and marginally. For a generalized linear mixed
model, conditional on the random effects, the regression parameters
have a
simple interpretation,
such as differences in means (linear regression), log-odds ratios (logistic
regression), or log rate ratios (Poisson regression). Often, though,
one is
also interested in the effects of the covariates on the
population-averaged or marginal mean, obtained by integrating the conditional
mean over the distribution of the
random effects. However, there is typically no closed form expression
for the
marginal mean as a function of the covariates. As such, there is no simple
expression for the marginal model. For example, for a binary outcome,
we would
want to formulate a table of the odds ratios for a one unit increase in each
covariate, given the other covariate values. The typical generalized
linear mixed (logistic regression) model with normal random effects
does not
provide a simple expression for the marginal odds ratios.

As an alternative to the marginalized random effects models of Heagerty
(\citeyear{Heagerty1999}),
but restricted to the setting of clustered binary data, Wang and Louis
(\citeyear{Wang2003})
proposed a random intercept generalized linear mixed model in which
both the
conditional model (conditional on the random effect) and the marginal model
(integrated over the distribution of the random intercept) follow a logistic
regression model, with model parameters proportional to each other. The random
intercept in the model of Wang and Louis (\citeyear{Wang2003}) follows
a ``bridge'' distribution.
The results of Wang and Louis (\citeyear{Wang2003}) hold for a model
with only a single random
intercept for all responses within a cluster. The restriction to models with
only a random intercept is somewhat unappealing for longitudinal
studies, as the
degree of association among a pair of repeated measures from two different
time points typically depends on their time separation. To take the declining
correlation into account, one could extend the model to have a random intercept
plus a random slope with time, where the random intercept and slope
follow a
bridge distribution. Unfortunately, a linear combination of random variables
from the bridge distribution no longer follows a bridge distribution,
so that
the desired property that the marginal model is of logistic form no longer
holds.

%----
In this paper we propose a modification of the bridge random intercept
model to
handle longitudinal data. In particular, we propose separate, but correlated,
random intercepts at each occasion. A multivariate density using a
copula model
for the random intercepts from different time points assures that the marginal
density of each random effect follows a bridge distribution. The
proposed model
allows for a flexible marginal correlation among the repeated binary outcomes,
including a declining association with increasing time separation while
retaining the property that the marginal probabilities follow a logistic
regression model. Further, the within-subject association has an appealing
interpretation in
terms of Kendall's $\tau$ between pairs of random intercepts as well as
Kendall's $\tau$ between any pair of repeated responses. The proposed model
can also be thought of as a modification of the correlated random normal
intercepts generalized linear mixed model for longitudinal binary
proposed by
Albert et al. (\citeyear{Albert2002}); however, the marginal model of
Albert et al. (\citeyear{Albert2002}) is not
logistic. The proposed model is more analogous to probit-normal
marginal models
for longitudinal binary data [Caffo, An and Rohde (\citeyear{Caffo2007});
Caffo and Griswold (\citeyear{Caffo2006})].

Except for the linear mixed model, there is typically no closed form
expression for the marginal likelihood (integrated over all possible
values of
the random effects) for any generalized linear mixed model. Thus, numerical
integration techniques must be used to approximate the likelihood, including
the likelihood based on our proposed approach here. These numerical
integration techniques include Laplace approximations, and Gauss-Hermite
quadrature, and Monte Carlo integration algorithms. Poor numerical
approximations to the likelihood will lead to biased estimates for the fixed
effects and variance components. Pinheiro and Bates (\citeyear
{Pinheiro1995}) showed that their
Monte Carlo importance sampling algorithm had good properties, and it
has been
implemented in standard generalized linear mixed models software, including
PROC NLMIXED in SAS or the NLME function in R.

The method-of-moments based generalized estimating equations (GEE) is an
alternative approach that can be used to estimate the marginal regression
parameters. Often, however, both the subject-specific conditional (on the
random effects) and the marginal regression parameters are of interest; with
GEE, only the latter are estimated. In addition, because GEE techniques [Liang
and Zeger (\citeyear{Liang1986}); Fitzmaurice, Laird and Rotnitzky (\citeyear
{F1993}); Diggle et al. (\citeyear{D2002})] for
estimation of
marginal regression parameters are not likelihood based, these methods
cannot be
used for prediction of the joint probability of the responses over
time. For
making inferences about the regression parameters, likelihood ratio
tests are
not available for hypotheses testing, and likelihood based model diagnostics
cannot be used with the GEE approach. Although beyond the scope of this paper,
with missing data, a full likelihood method typically gives less bias
than GEE
methods; the latter require the restrictive assumption that outcomes are
missing completely at random. Lee and Nelder (\citeyear{Lee2004})
document the drawbacks of
GEE methods even in cases when the main interest lies only in the marginal
regression parameters.

%s2 ###
\section{Random effects model with a bridge random effects distribution}

Although longitudinal data are clustered, there is in addition an implicit
ordering of the repeated measures on each subject. For ease of
presentation, we
assume that $n$ independent subjects are observed at a common set of
$t=1,\ldots,m$
times. Note, the model and associated methodology can be used when the
observation times $t_1<\cdots<t_{m_i}$ are unequally spaced, and when
the grid
of observation times as well as number of observations $m_i$ vary from subject
to subject. The outcome at time $t$ is binary, that is, we let $Y_{it}
= 1$ if
subject $i$ has response $1$ (say, success) at time $t$, and $Y_{it} = 0$
otherwise. Each individual has a $J \times1$ covariate vector $x_{it},$
measured at time $t,$ which includes both time-stationary and time-varying
covariates. Our approach can be used with time-varying covariates, but
it is
assumed that the covariates are nonrandom; in particular, all
time-varying covariates are assumed to be \textit{external} covariates
in the sense
described by Kalbfleisch and Prentice (\citeyear{Kalbfleisch1980}).
Random time-varying covariates
can potentially lead to bias for any GLMM as described by Fitzmaurice
(\citeyear{Fitzmaurice1995}). We
are
primarily interested in making inference about the marginal
distribution of $
Y_{it},$ which is Bernoulli with probability $ p_{it}= p_{it}(\bfbeta) =
E(Y_{it} | {x}_{it}, \bfbeta) = \pr(Y_{it} = 1 | {x}_{it}, \beta)$
indexed by unknown parameter vector $\beta.$

Wang and Louis (\citeyear{Wang2003}) proposed the following random
intercept logistic
regression model for the conditional subject-specific probability
%
%e1 ###
%
\begin{equation}
p_{it} = p_{it}(b_i) = \pr( Y_{it}=1|b_i, x_{it},\bfbeta)
= \frac{\exp( b_i + \phi^{-1}x_{it}' \beta)}{1 + \exp( b_i + \phi
^{-1}x_{it}'
\beta)} ,
\label{cond}
\end{equation}
where, given the subject-specific random effect $b_i$, the $Y_{it}$'s
from the
same subject are assumed independent Bernoulli random variables, that
is, $ Y_{it}
| b_i \sim \operatorname{Bern}(p_{it})$. When $b_i$ follows a ``bridge'' distribution,
%
%e2 ###
%
\begin{equation}
f_b(b_i|\phi) = \frac{1}{2\pi} \frac{\sin(\phi\pi)}{\cosh(\phi
b_i) +
\cos(\phi
\pi)} \qquad
(-\infty< b_i < \infty) ,
\label{bridge}
\end{equation}
indexed by unknown parameter $\phi$ $(0 < \phi< 1 )$, the marginal
probability of success [Wang and Louis (\citeyear{Wang2003})] equals
%
%e3 ###
%
\begin{equation}
\pr[Y_{it}=1|x_{it},\beta]= E_b [ p_{it}(b_i) ] =\frac{\exp[ x_{it}'
\beta]}{1
+ \exp[ x_{it}' \beta]} ,
\label{marg}
\end{equation}
where $E_b$ denotes the expectation evaluated with respect to the
density of the
$b_i.$ Thus, the marginal probabilities follow a logistic regression model
similar to the conditional model given in (\ref{cond}), except with parameter
$\beta$ instead of parameter $\phi^{-1}\beta$. The bridge random
variable in (\ref{bridge}) has mean 0 and
$\phi$ is the rescaling parameter. In particular,
\[
\Var(b_i) = \frac{\pi^2}{3} \biggl( \frac{1}{\phi^2} -1 \biggr)
\]
so that the larger the value of $\phi, $ the smaller the variance. The bridge
distribution is symmetric about 0 and has heavier tails than the Gaussian
distribution but lighter tails than the Logistic distribution. It can
also be
shown to be a scale mixture of Gaussian random variables. The rescaling
parameter $\phi\in(0,1)$ can be interpreted as the attenuation
parameter that
controls attenuation of the marginal regression effect due to
integration of the
random effects [Neuhaus, Kalbfleisch and Hauck (\citeyear
{Neuhaus1991})]. For a random
effects logistic model, the
only disadvantage to the choice of the bridge over the normal density
for the
random effects is that the bridge is not the default for any packaged computer
programs. The bridge density has a closed form that is easily programmed,
although it still requires numerical integration to obtain the MLE.
Thus, the
computation necessary to obtain the MLE is on a par with other random effects
distributions (e.g., the normal), but the interpretability of the marginal
model parameters makes the bridge distribution an attractive choice.
For a more
in-depth description of properties of the bridge distribution, see Wang and
Louis (\citeyear{Wang2003,Wang2004}).

Here, we propose a model with distinct, but correlated, random bridge
intercepts at each time point, that is, $b_i$ in (\ref{cond}) is
replaced by a
separate random intercept at time $t,$ say, $b_{it},$ where each
$b_{it}$ follows
a bridge distribution and the $b_{it}$'s from the same subject have a flexible
association structure. Specifically, we now let $\mathbf
{b}_i=(b_{i1},\ldots,b_{im})$ denote
the vector of random intercepts at the $m$ time points for subject
$i.$ Given the vector of random effects $\mathbf{b}_i,$ the $Y_{it}$'s
for subject $i$ are assumed to be independent Bernoulli random
variables, that is, $ Y_{it} | \mathbf{b}_i \sim \operatorname{Bern}(p_{it})$, where
%
%e4 ###
%
\begin{equation}
p_{it}
=
\frac{\exp( b_{it} + \phi^{-1}x_{it}' \beta)}
{1 + \exp( b_{it} + \phi^{-1}x_{it}'\beta)} ,
\label{cond3}
\end{equation}
and the $(m\times1)$-dimensional $\mathbf{b}_i$ has a multivariate
density such that
the marginal density of each $b_{it}$ is a bridge distribution as in
(\ref{bridge}). For simplicity, we assume the parameter $\phi$ of the bridge
distribution is the same for all times. Since $b_{it}$ has a bridge
distribution, the marginal success probability will be of the logistic
form in~(\ref{marg}). For the purpose of building a flexible
association among
$\mathbf{b}_i$, as well as assuring the desired marginal density of
each $b_{it}$, we use
a Gaussian copula [Nelsen (\citeyear{Nelson1999})] for $\mathbf{b}_i.$
Mathematically, a
copula is a simple way of formulating an m-dimensional multivariate
distribution,
and is specified as a function of the marginal CDF's. If
$F_1(w_1),F_2(w_2),\ldots,F_m(w_m)$ are the cumulative distribution
functions of
the random variables $W_1,W_2,\ldots,W_m,$ respectively, then there
exists a
function $C$ such that the joint CDF is
$F(w_1,\ldots,w_m)=C(F_1(w_1),\ldots,F_m(w_m)),$ with one-dimensional marginal
distributions given by $F_1(w_1),\ldots,F_m(w_m)$. The concept and application
of copulas are illustrated in Nelsen (\citeyear{Nelson1999}) and Joe~(\citeyear{Joe1997}).

To formulate the Gaussian copula for $\mathbf{b}_i,$ we form a $m
\times1$ vector,
$Z_i=[Z_{i1},\ldots,Z_{im}]',$ which is multivariate normal with mean
vector $0$
and covariance matrix $\Sigma,$ where the diagonal elements of $\Sigma$
equal 1
so that $\Sigma$ is also the correlation matrix. Note, for identifiability,
we restrict $\Var(Z_{it})$ to equal 1; if $\Var(Z_{it})$ is left as a
parameter
to estimate, then $\Var(b_{it})$ would be a function of both $\phi$ and
$\Var(Z_{it}),$ but only one of the two would be estimable. We let
$\rho_{ist}
= \Corr(Z_{is},Z_{it})$ denote the correlation between $Z_{is}$ and $Z_{it};$
various choices for the structures of $\rho_{ist}$ are discussed
below. Using
the probability integral transform [Hoel, Port and Stone (\citeyear
{Hoel1971})],
$b_{it}=F_{b}^{-1}(\Phi(Z_{it}))$ has CDF $F_{b}(b_{it}),$ where $\Phi$
is the
CDF of a standard normal density,
\[
F_{b}^{-1}(u) = \frac{1}{\phi} \log\biggl[ \frac{ \sin(\phi\pi u)}
{\sin\{ \phi\pi(1 - u) \} }
\biggr]
\]
is the inverse cumulative distribution function of
$b_{it}$ for $0 < u_{it} < 1$, and
%
%e5 ###
%
\begin{equation}
F_{b}(b_{it}) = 1 - \frac{1}{\pi\phi} \biggl[ \frac{\pi}{2} -
\arctan\biggl\{ \frac{\exp(\phi b_{it}) + \cos(\phi\pi)}
{\sin(\phi\pi)}
\biggr\}
\biggr] \label{br2}
\end{equation}
denotes the cumulative distribution function of the bridge
distribution. Thus, $b_{it} = F_{b}^{-1}(\Phi(Z_{it}))$ has the
marginal bridge distribution of interest, and the $b_{it}$'s within
a subject are correlated due to the correlation among the $Z_{it}$'s.

To fully specify the distribution of $Z_i=[Z_{i1},\ldots,Z_{im}]',$ we must
specify the correlation matrix $\Sigma.$ A popular longitudinal
correlation structure is the autoregressive(1) $\operatorname
{AR}(1)$ structure,
%
%e6 ###
%
\begin{equation}
\rho_{ist} = \Corr(Z_{is},Z_{it}) = \rho^{|t-s| } ,
\label{ar1}
\end{equation}
where $-1 < \rho< 1.$ In principle, any suitable longitudinal correlation
structure for the $Z_{it}$'s could be assumed, such as Toeplitz,
ante-dependence, or anisotropic exponential. Alternatively, as
discussed by
Hougaard (\citeyear{Hougaard2000}), Kendall's $\tau$ is often
recommended as a measure of
association between a pair of continuous random variables since it is invariant
to monotone transformations of the random variables. For a pair of normal
random variables, Hougaard (\citeyear{Hougaard2000}) shows that
Kendall's $\tau$ equals
%
%e7 ###
%
\begin{equation}
\tau_{ist} = \frac{2\arcsin(\rho_{ist})}{\pi} ,
\label{tau_ijk}
\end{equation}
where $\arcsin(\cdot)$ is the inverse sin function and $-1 \leq
\tau_{ist} \leq1.$ Because the bridge random variables $b_{is}$ and
$b_{it}$ are monotone transformations of $Z_{is}$ and $Z_{it},$ and
Kendall's $\tau$ is invariant to monotone transformations, then
(\ref{tau_ijk}) is also Kendall's $\tau$ between the bridge random
variables $b_{is}$ and $b_{it}.$ This is important because
(\ref{tau_ijk}) is easy to calculate and it shows that the copula
model can capture the full range of possible association between
$b_{is}$ and $b_{it}$. One possibility we suggest is specifying the
association model in terms of $\tau_{ist},$ such as $\operatorname
{AR}(1)$,% -
%
%e8 ###
%
\begin{equation}
\tau_{ist} = \tau^{|t-s|},
\label{tau_ar1}
\end{equation}
and then transforming back to $ \rho_{ist} = \sin(\pi\tau_{ist}/2) $
to get the multivariate normal correlation matrix $\bfSigma.$ The
relationship between the Kendall's $\tau$ for $b_{is}$ and $b_{it}$ and
the Kendall's $\tau$ for $Y_{is}$ and $Y_{it}$ can only be computed
numerically.\looseness=1

%f1 ###
%
\begin{figure}[b]

\includegraphics{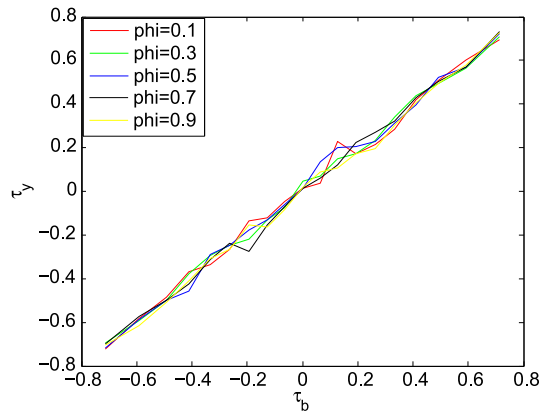}

\caption{Plot of Kendall's $\tau$ for
$(Y_{is},Y_{it})$ (denoted $\tau_Y$) versus Kendall's
$\tau$ for $(b_{is},b_{it})$ (denoted $\tau_B$).}
\label{fig1}
\end{figure}

To explore the extent of the associations that the bridge random
effects can
induce, we considered a plot of the relationship between Kendall's
$\tau
$ for
$(b_{is},b_{it})$ and Kendall's $\tau$ for $(Y_{is},Y_{it}),$
calculated via
Monte Carlo simulation (see Figure~\ref{fig1}). For this illustration, we
considered two
time points with bridge model
\[
\pr( Y_{it}=1|b_i, x_{it},\bfbeta) =
\frac{\exp[ b_i + (3 - 2t)\phi^{-1}]}
{1 + \exp[ b_i + (3 - 2t)\phi^{-1}]}
\]
for $t=1,2,$ and let $\phi=0.1, 0.3, 0.5, 0.7, 0.9.$ From Figure~\ref
{fig1} we see
that the curves follow closely along the 45 degree line, meaning that
Kendall's $\tau$ for $(b_{is},b_{it})$ is a close approximation to Kendall's
$\tau$ for $(Y_{is},Y_{it}).$ Further, in terms of Kendall's $\tau$, the
range of association is $-$1 to 1, and there are no constraints on the
association. We have found that this is not true for the usual correlation
coefficient, that is, $\Corr(b_{is},b_{it})$ can be much different than
$\Corr(Y_{is},Y_{it})$.

Here, we briefly discuss identifiability issues, which are similar to
identifiability issues for a linear mixed model. With both $\phi$ and
$\rho_{ist}$ in the model, identifiability issues can arise, depending
on the
number of pairs of time points, and the model for the association over time.
When there are only $m=2$ times, the model is not identified if both
$\rho_{ist}$ and $\phi$ are left unspecified, that is, with only two
times points,
the association between $b_{is}$ and $b_{it}$ is completely determined
by either
the variance of the random effects (a function of $\phi,$) or the correlation
between random effects (a function of $\rho_{ist}$), but not both. As
is the
case for a linear mixed model, for three or more repeated measures, the
identifiability of the model will depend on the specified correlation structure.
For example, for three time points, there are three pairs of times, so that
we could have $\phi$ in the model, as well as a model for $\rho_{ist}$
that has
two parameters. The above identifiability issues do not arise when one models
$\rho_{ist}$ and/or $\phi$ as a function of cluster-level (time-stationary)
covariates, although identifiability issues could arise, as in any regression
model, if one models $\rho_{ist}$ and/or $\phi$ as a function of too many
cluster-level covariates.

The maximum likelihood estimates for the marginal likelihood,
integrated over
the random effects, say,
\[
L(\beta,\phi,\rho) =
\prod_{i=1}^n \int_{b_i}
\Biggl[
\prod_{t=1}^m p_{it}^{y_{it}} ( 1-p_{it} )^{(1-y_{it})} \Biggr]
f(\mathbf{b}_i|\phi,\rho)\,d\mathbf{b}_i
\]
can be obtained using a simulation maximization method such as the
Monte Carlo
importance sampling algorithm described by Pinheiro and Bates
(\citeyear
{Pinheiro1995}), and
implemented in PROC NLMIXED in SAS (V9.2) or the NLME function in R;
the estimated covariance matrix is obtained using the Pinheiro and
Bates (\citeyear{Pinheiro1995})
numerical approximation to the inverse of the negative second derivative
(information) matrix. A SAS macro for fitting the model is available
upon request from the first author. If there are missing outcome data that
are \textit{missing at random} [Rubin (\citeyear{Rubin1976}); Laird
(\citeyear{Laird1988})], each individual
contributes $m_i \leq m$ conditionally independent (given the random effects)
Bernoulli random variables with success probabilities given by (\ref
{cond3}) to
the overall likelihood, and the marginal likelihood is again formed by
integrating over the random effects. Appealing to large sample theory for
generalized linear mixed models [Fahrmeir and Tutz (\citeyear
{Fahrmeir2001})], if the likelihood is
correctly specified, the maximum likelihood estimates are consistent,
asymptotically normal, and the large sample variance of the maximum likelihood
estimates can be consistently estimated by the inverse of the negative second
derivative (information) matrix.

In order for the Monte Carlo importance sampling algorithm of Pinheiro
and Bates
(\citeyear{Pinheiro1995}) to provide a computationally stable and
efficient way of approximating
the marginal likelihood, one must carefully choose the importance sampling
distribution from which to sample. We have found that the Pinheiro and Bates
(\citeyear{Pinheiro1995}) suggestion of a multivariate normal
approximation for
$ [
\prod_{t=1}^m p_{it}^{y_{it}} ( 1-p_{it} )^{(1-y_{it})} ]
f(\mathbf{b}_i|\phi,\rho)$
produces stable results. Further, once the likelihood is approximated, we
suggest using the Newton--Raphson algorithm to obtain the maximum likelihood
estimate, which requires good starting values for the parameter
estimates. We
have found that using the ordinary logistic regression estimates of
$\bfbeta$ as
the starting values leads to computational stability. In the present study
(discussed in the next section), with seven time points, the algorithm
is stable
and converged quite fast (within 2 minutes). In general, an increase in the
dimension of the integration has both positive and negative trade-offs. First,
with an increase in the number of time points (or dimension of the integration),
there is more information from which to estimate the association parameters
$\phi$ and $\tau$ (or $\rho$), so that the chances of a flat or multimodal
likelihood is far less than it might be with fewer time points.
However, with
an increase in the dimension of the random effects, the computation
required to
maximize the likelihood increases. Similar to the approach recommended by
Albert et al. (\citeyear{Albert2002}), we suggest performing at most 50
iterations of the
Newton--Raphson algorithm, with 50 Monte Carlo samples drawn for iterations
1--19, 100 Monte Carlo samples drawn for iterations 20--39, and 1000 iterations
for iterations 40--50.\looseness=1

%s3 ###
\section{Example: Longitudinal study of cardiac function
in children born to women infected with HIV-1}

In this section we illustrate the application of the proposed
methodology to
the analysis of the data from children born to women infected with HIV-1
described in the \hyperref[intro]{Introduction}. In the P$^2$C$^2$ study, a birth cohort
of 401
infants born to women infected with HIV-1 were to have cardiovascular function
measured approximately every year from birth to age 6, giving up to 7
measurements on each child. Of these 401 infants, 74 $(18.8\%)$ were HIV
positive, and 319 $(81.2\%)$ were HIV negative. The main scientific
interest is
in determining if HIV-1 infected children are more likely to have abnormal
``pumping ability of the heart'' at time $t$ (1${}={}$yes, 0${}={}$no).
The main covariate
of interest is the effect of HIV infection in the child; other
covariates that
could be potential confounders are mother's smoking status during pregnancy
(1${}={}$yes, 0${}={}$no), gestational age (in weeks) and birth-weight
standardized
for age
(1${}={}$abnormal, 0${}={}$normal). A~child of a mother who smokes is
expected to have
worse heart function. Children with younger gestational age and lower
birth-weight (standardized for gestational age) may also be at risk for cardiac
problems.

Thus, to examine the effect of HIV infection in the infants, we
considered the
following marginal logistic regression model,
%
%e9 ###
%
\begin{eqnarray}
\log\biggl[ \frac{p_{it}}{ 1-p_{it} } \biggr] &=& b_{it} +
\beta_0 + \beta_1 t + \beta_2 \HIV_i + \beta_{12} t * \HIV
_i\nonumber
\\[-8pt]
\\[-8pt] &&{}+\beta_3
\smoke_i
+ \beta_4 \age_i + \beta_5 \wt_i
\label{log_reg}
\nonumber
\end{eqnarray}
for $t=0,1,\ldots,6,$
where $\HIV_i$ equals 1 if the $i${th} child is born with HIV-1 and
equals 0
if otherwise; $\smoke_i$ equals 1 if the mother smoked during
pregnancy, and 0
otherwise; $\age_i$ is the gestational age (in weeks); and $\wt_i$
equals 1
if the child's birth-weight for gestational age was abnormal, and 0 otherwise.

Here, we compare our proposed estimation technique with four alternative
approaches:
\begin{longlist}
\item[(1)] the bridge random effects model of Wang and Louis
(\citeyear{Wang2003}) with a
single bridge random effect;
\item[(2)] Heagerty's (\citeyear{Heagerty1999})
\textit{marginalized} random
effects model with a linear term for time in the random effects
variance, as
implemented using the R-macro:
\begin{center}
\url{http://faculty.washington.edu/heagerty/Software/LDA/};
\end{center}
\item[(3)] the maximum likelihood estimates assuming a parametric Bahadur
representation
of the multinomial distribution [Bahadur (\citeyear{Bahadur1961})] with
an $\operatorname{AR}(1)$ correlation
structure between $Y_{is}$ and $Y_{it},$ that is,
%
%e10 ###
%
\begin{equation}
\Corr(Y_{is},Y_{it}) = \Gamma^{|t-s|}; \label{AR1c}
\end{equation}
\item[(4)]
generalized estimating equations (GEE) with an $\operatorname{AR}(1)$
correlation
structure for
$\Corr(Y_{is},Y_{it}).$ For the proposed approach, we use two
association models
for the bridge random intercepts, one is $\operatorname{AR}(1)$ on
the $\Corr(b_{is},b_{it}),$
and the other is $\operatorname{AR}(1)$ on the Kendall's $\tau$
between $b_{is}$ and $b_{it}.$
All
approaches assume the same marginal model, but different association
structures. With the exception of the random effects model with a
single bridge
random effect, the association between pairs of outcomes decreases as
the time
separation increases.
\end{longlist}

Because the
Bahadur representation is used, we briefly describe it here. In the
Bahadur distribution, the marginal model is $p_{it}$ in (\ref{marg}). Next,
we define the standardized variable $S_{it}$ to be
\[
S_{it} = \frac{ Y_{it} - p_{it} } { \{ p_{it}(1-p_{it}) \}^{1/2} }.
\]
The pairwise correlation between $Y_{is}$ and $Y_{it}$ is
$\Gamma_{st} = E(S_{is}S_{it}),$
and the $M${th}-order correlation between the first $M$ responses is defined
as $\Gamma_{12\ldots M} = E(S_{i1}S_{i2} \cdots S_{iM}).$ The $M${th}-order
correlation between any $M$ of the $m$ repeated binary responses is defined
similarly. Then the Bahadur representation of the $2^m-1$ multinomial
probabilities corresponding to the joint distribution of
$(Y_{i1},Y_{i2},\ldots,Y_{im})$ is
\begin{eqnarray}\label{bah}
&&\pr\{(Y_{i1}=y_1), (Y_{i2}=y_2), \ldots, (Y_{im}=y_m)|X_i,\beta,
\Gamma
\}
\nonumber
\\
&& \qquad=\Biggl\{ {\prod_{t=1}^{m}} p_{it}^{y_{it}}(1
-p_{it})^{1-y_{it}} \Biggr\}\\
&& \qquad\quad{}\times
\biggl\{
1 + {\sum_{s t}} \Gamma_{st} s_{is}s_{it} +
{\sum_{s t u}} \Gamma_{stu} s_{is}s_{it}s_{iu} + \cdots+
\Gamma_{1\ldots m} s_{i1} \cdots s_{im} \biggr\} .
\nonumber
\end{eqnarray}
In obtaining the MLE from the Bahadur representation, we
assumed all
fifth and higher correlations are 0 $(\Gamma_{stuvw}= \cdots= \Gamma
_{1\ldots m} =
0);$ we assumed all fourth-order correlations are the same, regardless
of the
sets of times ($\Gamma_{stuv}= \Gamma_{s't'u'v'}$ for all $stuv \neq
s't'u'v'$);
and we assumed all third-order correlations are the same, regardless of the
sets of times ($\Gamma_{stu}= \Gamma_{s't'u'}$ for all $stu \neq
s't'u'$). The
model for the pairwise correlations $\Gamma_{st}$ is $\operatorname
{AR}(1)$ as in (\ref{AR1c}).

The importance sampling algorithm of Pinheiro and Bates (\citeyear
{Pinheiro1995}) was used to
calculate the MLE for the bridge random effects model, with the same starting
seed and the same number of Monte Carlo draws (400) for each model.
Performing a sensitivity analysis, we found very little difference in the
estimates and standard errors with 100, 200, 300, or 400 Monte Carlo draws.
To obtain the estimates, we wrote a SAS macro using PROC NLMIXED; the macro
can be obtained from the first author. For the model with a single bridge
random effect, the SAS macro takes approximately 30 seconds to
calculate the
estimates on a Dual Core, 2.7~GHz, 4~GB Ram computer; for either the
$\operatorname{AR}(1)$ model
on the correlation or Kendall's~$\tau$, the SAS macro takes
approximately 2
minutes to calculate the estimates.

%t2 ###
%
\begin{table}
\tabcolsep=0pt
\caption{Comparison of parameter estimates under alternative
models for
the within-subject
association}\label{tab2}
\begin{tabular*}{\textwidth}{@{\extracolsep{\fill}}lld{2.3}cd{2.2}d{2.3}@{}}
\hline
\textbf{Effect} & \multicolumn{1}{c}{\textbf{Model}} &\multicolumn
{1}{c}{\textbf{Estimate}}
& \multicolumn{1}{c}{$\bolds{\mathit{SE}}$} & \multicolumn
{1}{c}{\textbf{$\bolds Z$-statistic}} & \multicolumn{1}{c@{}}{\textbf
{$\bolds p$-value}}
\\
\hline
Intercept & Bridge & 1.827 & 1.459 & 1.25 & 0.211 \\
& $\operatorname{AR}(1)$-corr & 1.374 & 1.590 & 0.86 & 0.388 \\
& $\operatorname{AR}(1)$-$\tau$ & 1.389 & 1.514 & 0.92 & 0.359 \\
& Heagerty & 2.073 & 1.407 & 1.47 & 0.141\\
& Bahadur & 1.763 & 1.352 & 1.30 & 0.193 \\
& GEE & 1.959 & 1.506 & 1.30 & 0.193 \\
Time & Bridge & -0.641 & 0.080 & -8.02 & {<}0.001 \\
& $\operatorname{AR}(1)$-corr & -0.812 & 0.102 & -7.98 & {<}0.001 \\
& $\operatorname{AR}(1)$-$\tau$ & -0.815 & 0.094 & -8.67 & {<}0.001
\\
& Heagerty & -0.612 & 0.063 & -9.64 & {<}0.001 \\
& Bahadur & -0.637 & 0.088 & -7.28 & {<}0.001\\
& GEE & -0.642 & 0.098 & -6.57 & {<}0.001\\
HIV & Bridge & -0.075 & 0.266 & -0.28 & 0.777 \\
& $\operatorname{AR}(1)$-corr & -0.082 & 0.264 & -0.31 & 0.756 \\
& $\operatorname{AR}(1)$-$\tau$ & -0.076 & 0.259 & -0.29 & 0.769 \\
& Heagerty & -0.038 & 0.249 & -0.15 & 0.879 \\
& Bahadur & -0.037 & 0.269 & -0.14 & 0.891 \\
& GEE & -0.073 & 0.264 & -0.28 & 0.782 \\
TIME${}\times{}$HIV & Bridge & 0.234 & 0.135 & 1.73 & 0.084 \\
& $\operatorname{AR}(1)$-corr & 0.336 & 0.170 & 1.97 & 0.049 \\
& $\operatorname{AR}(1)$-$\tau$ & 0.323 & 0.160 & 2.02 & 0.044 \\
& Heagerty & 0.226 & 0.101 & 2.23 & 0.025 \\
& Bahadur & 0.213 & 0.140 & 1.53 & 0.128 \\
& GEE & 0.251 & 0.156 & 1.61 & 0.108 \\
MOM SMOKE & Bridge & -0.182 & 0.176 & -1.03 & 0.303 \\
& $\operatorname{AR}(1)$-corr & -0.170 & 0.185 & -0.92 & 0.359 \\
& $\operatorname{AR}(1)$-$\tau$ & -0.179 & 0.177 & -1.01 & 0.314 \\
& Heagerty & -0.197 & 0.187 & -1.05 & 0.292\\
& Bahadur & -0.206 & 0.172 & -1.20 & 0.231 \\
& GEE & -0.200 & 0.173 & -1.15 & 0.248 \\
GEST AGE & Bridge & -0.045 & 0.037 & -1.22 & 0.225 \\
& $\operatorname{AR}(1)$-corr & -0.038 & 0.040 & -0.93 & 0.352 \\
& $\operatorname{AR}(1)$-$\tau$ & -0.037 & 0.038 & -0.95 & 0.341 \\
&Heagerty & -0.052 & 0.036 & -1.45 & 0.149\\
& Bahadur & -0.043 & 0.034 & -1.26 & 0.207 \\
& GEE & -0.048 & 0.038 & -1.26 & 0.207 \\
Low birth Wt & Bridge & 0.086 & 0.190 & 0.45 & 0.652 \\
& $\operatorname{AR}(1)$-corr & 0.122 & 0.198 & 0.62 & 0.536 \\
& $\operatorname{AR}(1)$-$\tau$ & 0.136 & 0.191 & 0.71 & 0.477 \\
& Heagerty & 0.078 & 0.191 & 0.41 & 0.683\\
& Bahadur & 0.096 & 0.173 & 0.55 & 0.581 \\
& GEE & 0.083 & 0.193 & 0.43 & 0.667\\
\hline
\end{tabular*}
\end{table}
\setcounter{table}{1}
%t2 ###
%
\begin{table}
\tabcolsep=0pt
\caption{(Continued)}
\begin{tabular*}{\textwidth}{@{\extracolsep{\fill}}llcc@{}}
\hline
\textbf{Parameter} & \textbf{Model} & \multicolumn{1}{c}{\textbf{Estimate}} &
{$\bolds{95\%}$ \textbf{confidence interval}}\\
\hline
{$\phi$} & Bridge & 0.847 &
{[0.788, 0.906]} \\
& $\operatorname{AR}(1)$-corr & 0.686 &
{[0.556, 0.815]} \\
& $\operatorname{AR}(1)$-$\tau$ & 0.731 &
{[0.634, 0.827]} \\[3pt]
{$\rho$} & $\operatorname{AR}(1)$-corr & 0.841
&
{[0.725, 0.957]} \\
{$\tau$} & $\operatorname{AR}(1)$-$\tau$ &
0.749
&
{[0.651, 0.847]}\\[3pt]
{$\Gamma$} & Bahadur $\operatorname{AR}(1)$ & 0.206
& {[0.107, 0.304]} \\
\hline
\end{tabular*}
\end{table}

Table~\ref{tab2} gives the estimates of $\beta$ obtained using the
different approaches.
We see that the results are generally similar. Although well within sampling
random error, if one chooses a 0.05 level of significance as a cutoff, the
parameter of greatest scientific interest, the interaction between Time
and HIV
status, is significant using Heagerty's approach as well as our proposed
approach with an $\operatorname{AR}(1)$ model for $\rho$ or
Kendall's $\tau$, but not
using the
single bridge random intercept model, GEE, or the Bahadur representation.
With a significant interaction, the odds ratio for children with HIV versus
those without HIV increases over time. For example, using results from the
bridge model with $\operatorname{AR}(1)$-$\tau$, children with HIV
have $\exp(\widehat
{\beta}_2
+
\widehat{\beta}_{12} t) = \exp(-0.076 + 0.323 t)$ times the odds of
having an
abnormal pumping ability than children without HIV at time $t.$ Thus,
at 6
years of age, children with HIV have approximately 6 times the odds (or
$e^{-0.076 + 0.323 \times6} = 6.4$) of having an abnormal pumping ability.

For the main parameter of interest, it appears from Table~\ref{tab2}
that Heagerty's
approach yields a discernibly smaller standard error estimate for the
interaction term; however, we caution that this result cannot be
expected in
general. Overall, there is no clear pattern for the magnitudes of standard
errors from one approach versus another. The $\operatorname{AR}(1)$
associations from
the bridge
random effects models can be interpreted as follows. Random intercepts
that are
1 year apart have a correlation estimated to be 0.84 and Kendall's
$\tau$
estimated to be 0.75; both estimates indicate a high correlation among the
repeated binary responses. To compare the fit of the bridge models, one can
examine the Akaike information criterion (AIC) for the models, where smaller
AIC is defined as better. The AIC for the $\operatorname{AR}(1)$
model based on $\rho$ is
5522.6 and for the model based on $\tau$ is 5520.6. The AIC for the Bridge
model of Wang and Louis (\citeyear{Wang2003}) is 5534.2. This suggests
that the $\operatorname{AR}(1)$ model
based on $\tau$ provides a slightly better fit than the
$\operatorname{AR}(1)$ model
based on
$\rho;$ both provide better fits than the bridge random effects model
with a
single random effect. For all practical purposes, the fits of the two models
are almost indistinguishable. Thus, either model is appropriate for
these data
and the choice between them can be made in terms of ease of
interpretation of
the $\operatorname{AR}(1)$ association parameter.

%s4 ###
\section{Simulation study}

We conducted a simulation study to explore the finite sample properties
of the
proposed bridge generalized linear mixed models. Specifically, we
compared the
ML estimator for the bridge random effects model, the ML estimator
assuming a
Bahadur distribution, and a GEE estimator of $\beta$. To ensure feasibility
of the simulation study, we restricted the number of occasions to $m=3$ and
considered a simple two-group ($50\dvtx50$ mixture) study design configuration
(e.g., active treatment versus placebo), with 50 subjects in each
group. We
simulated from two ``true'' models: (1) a generalized linear mixed model
with a
bridge distribution, and (2) the Bahadur representation.

Let $x_i=0,1$ indicate group membership, and $Y_{it}$ again denote the binary
outcome at time $t,$ $t=1,2,3$. When simulating from the bridge or Bahadur
models, we let the true marginal logistic model
be
\[
\operatorname{logit}(\pr[Y_{it}=1|x_{it}]) = \beta_{0} + \beta_{x}
x_{i} + \beta
_{\tau}
t ,
\]
with $\beta_{0} = -1.0$, $\beta_\tau=-0.5$, and $\beta_x=1.0.$
For the bridge random effects model, we specified an $\operatorname
{AR}(1)$ model for the
correlation structure for the $Z_{it}$'s in (\ref{ar1}), that is,
\[
\rho_{ist} = \Corr(Z_{is},Z_{it}) = \rho^{|t-s|}
\]
for three possible true values of $\rho=0.1,0.3,0.6,$ and we also let
$\Var(Z_{is})=1$. For the Bahadur representation given in (\ref
{bah}), we
specified an $\operatorname{AR}(1)$ model for the correlation
structure for the $Y_{it}$'s,
\[
\Gamma_{ist} = \Corr(Y_{is},Y_{it}) = \Gamma^{|t-s|}
\]
for three possible true values of $\Gamma=0.1,0.25,0.4;$ we set
$\Gamma_{123}=0.$ The constraints for the Bahadur representation did
not allow
$\Gamma> 0.4.$ For each simulation configuration, 1000
simulation replications were performed. Our simulations were performed using
PROC NLMIXED in SAS, with 200 Monte Carlo draws.

For each simulation replication we estimated the $\beta$'s by fitting
the bridge
random effects model with an $\operatorname{AR}(1)$ structure on the
underlying
$Z_{is}$'s, a
Bahadur model with an $\operatorname{AR}(1)$ structure on the
$Y_{is}$'s, and GEE with
an $\operatorname{AR}(1)$
structure on the $Y_{is}$'s. Note that the GEE will be asymptotically unbiased
when data are simulated from either a bridge random effects model or a Bahadur
distribution. The MLE from the bridge random effects model will be
asymptotically unbiased when data are simulated from a bridge random effects
model, but could be biased when data are simulated from the Bahadur. Similarly,
the MLE when assuming a Bahadur distribution will be asymptotically unbiased
when data are simulated from the Bahadur representation, but could be biased
when data are simulated from a bridge random effects model. The purpose
of the
simulation was to explore the robustness of the MLE from the bridge random
effects model under mis-specification of the likelihood. We explored the
properties of the three estimators with respect to bias, mean square error
(MSE), and coverage probability.

The results of the simulations reported in Table~\ref{tab3} indicate
that all of the
methods are approximately unbiased and have correct coverage probabilities,
even when the likelihood is misspecified. In general, the MLE from the
correctly specified likelihood tends to have the smallest MSE, although the
ratio of MSE's for pairs of approaches is at least $90\%$ for most
configurations. For example, the largest difference in ratios of MSE's when
simulating from the bridge random effects model is for $\rho=0.6$ and
$\beta_x=1;$ in this case, the ratio of the bridge MSE to the Bahadur
MSE is
$90.4\%,$ which suggests the Bahadur MLE is $90\%$ efficient in this case.

%t3 ###
%
\begin{sidewaystable}
\tabcolsep=0pt
\caption{Results of simulation study.
The true marginal logistic model
has parameters $(\beta_\tau,\beta_x) = (-0.5,1.0)$}
\label{tab3}
\begin{tabular*}{\textwidth}{@{\extracolsep{\fill
}}llld{2.4}d{2.4}@{\hspace{3pt}}d{2.4}d{2.4}@{\hspace{3pt}}d{2.4}d{2.4}@{}}
\hline
\textbf{True} & & & & & & & & \\
\textbf{distribution} & & \multicolumn{1}{c}{\textbf{APPROACH}} &
\multicolumn{1}{c}{$\bolds{\beta_\tau=-0.5}$} & \multicolumn
{1}{c@{\hspace{3pt}}}{$\bolds{\beta_x=1.0}$}
&\multicolumn{1}{c}{$\bolds{\beta_\tau=-0.5}$} & \multicolumn
{1}{c@{\hspace{3pt}}}{$\bolds{\beta_x=1.0}$}
&\multicolumn{1}{c}{$\bolds{\beta_\tau=-0.5}$} & \multicolumn
{1}{c@{}}{$\bolds{\beta_x=1.0}$} \\
\hline
Bridge & & &
\multicolumn{2}{c}{$\rho=0.10$} &
\multicolumn{2}{c}{$\rho=0.30$} &
\multicolumn{2}{c@{}}{$\rho=0.60$} \\
& Simulation & Bridge ML & -0.505 & 1.001 & -0.509 & 1.009 & -0.507
& 1.012 \\
& average & Bahadur ML & -0.508 & 1.019 & -0.502 & 1.016 & -0.514 &
1.020 \\
& & GEE & -0.517 & 1.024 & -0.509 & 1.033 & -0.506 &
1.001 \\
\\
& Simulation & Bridge ML & 0.0291 & 0.0790 & 0.0297 & 0.0771 &
0.0282 &
0.0829 \\
& MSE & Bahadur ML & 0.0296 & 0.0793 & 0.0301 & 0.0782 & 0.0294 &
0.0917 \\
& & GEE & 0.0299 & 0.0823 & 0.0305 & 0.0842 & 0.0284 &
0.0851 \\
& Coverage & Bridge ML & 94.0 & 95.5 & 95.1 & 94.9 & 93.2 & 94.7 \\
& probability\tabnoteref[a]{tb3a} & Bahadur ML & 94.8 & 95.1 & 93.9 &
96.0 & 95.7 &
93.8
\\
& & GEE & 94.3 & 93.6 & 93.9 & 95.1 & 94.6 & 95.1
\\
Bahadur & & &
\multicolumn{2}{c}{$\Gamma=0.10$} &
\multicolumn{2}{c}{$\Gamma=0.30$} &
\multicolumn{2}{c@{}}{$\Gamma=0.60$} \\
& Simulation & Bridge ML & -0.510 & 1.027 & -0.508 & 1.001 & -0.518
& 0.966 \\
& average & Bahadur ML & -0.509 & 0.997 & -0.514 & 1.031 & -0.507 &
1.021 \\
& & GEE & -0.513 & 1.024 & -0.506 & 1.015 & -0.505 &
1.025 \\
\\
& Simulation & Bridge ML & 0.0299 & 0.0867 & 0.0278 & 0.1053 &
0.0241 &
0.1115 \\
& MSE & Bahadur ML & 0.0290 & 0.0809 & 0.0265 & 0.1036 & 0.0233 &
0.1113 \\
& & GEE & 0.0288 & 0.0888 & 0.0272 & 0.1057 & 0.0256 &
0.1366 \\
\\
& Coverage & Bridge ML & 93.4 & 95.2 & 93.0 & 94.7 & 93.2 & 94.7
\\
& probability\tabnoteref[a]{tb3a} & Bahadur ML & 94.4 & 95.7 & 95.2 &
95.1 & 92.8 &
94.8
\\
& & GEE & 95.5 & 95.4 & 95.9 & 95.0 & 93.8 & 93.6
\\
\hline
\end{tabular*}
\tabnotetext[a]{tb3a}{Coverage probability for a $95\%$ confidence
interval.}
\end{sidewaystable}

The results of this simulation study suggest that the MLE from the bridge
random effects model is approximately unbiased, and has correct coverage
probabilities, even when the likelihood is misspecified. We caution, however,
that when there are missing data and a misspecified likelihood, the MLE from
the bridge random effects model (and the GEE estimator and MLE from the Bahadur
model) could yield biased estimates.

%s5 ###
\section{Discussion}

In this paper we have proposed a correlated random intercepts model for
longitudinal binary data that leads to a marginal logistic regression model.
Although the main focus of this paper is on a marginal logistic model
for the
probability of response at each time point, the model also has the appealing
property that the probability of response at each time point,
conditional on the
random effect, is also of logistic form. Specifically, the logistic regression
parameters for the marginal and conditional models are proportional to each
other, with the proportionality factor determined by an ``attenuation
parameter.'' Thus,\vadjust{\goodbreak} the proposed approach can also be used if there is interest
in the conditional model. As discussed in the \hyperref[intro]{Introduction}, a variety of
generalized linear mixed models have previously been proposed that yield
logistic marginal models; however, none of them have the property that both
the marginal and conditional models are of logistic form. We note that the
proposed approach can be generalized to other link functions with an appropriate
bridge distribution, such as the complimentary log--log link for longitudinal
binary data with a positive stable random effect. Furthermore, the proposed
model can easily be fit using existing software, for example, PROC
NLMIXED in SAS.
For example, using the Gaussian copula, we can express the marginal likelihood
$L(\beta,\phi,\rho)$ in terms of standard nonlinear mixed-effects
models with
random effects $b_{it}.$ Then the model can be fit using SAS PROC NLMIXED,
the R function NLME, or any nonlinear mixed-effects software program
that is
flexible enough to allow transformations of the normal random effects.

Finally, the proposed method can be extended in a number of ways. First,
consider a joint longitudinal model for a binary and continuous outcome measured
over time. For a joint analysis of both outcomes, the longitudinal
binary data
can be modeled as in Section 3 and the continuous outcome can be
modeled using a
standard linear mixed effects model. Correlation between the
longitudinal binary
and continuous outcomes can be induced by specifying correlations
between the
random effects in the linear mixed effects model for the continuous
outcomes and
the bridge random effects in the model for the longitudinal binary outcomes.
The second potential extension applies to the problem of
``informative'' dropout,
with the probability of dropout related to possibly unobserved
outcomes. One
approach for handling informative dropout is to model the (continuous) dropout
time process with a parametric frailty model [Hougaard (\citeyear
{Hougaard2000})], in which the
frailty is correlated with the bridge random effects in the model for the
longitudinal binary outcomes.

%suskaldyti doi

\printaddresses


\begin{thebibliography}{99}
%b1 ###
\bibitem[\protect\citeauthoryear{}{2002}]{Albert2002}
\textsc{Albert, P.~S.}, \textsc{Follmann, D.~A.}, \textsc{Wang,
S.~A.} and \textsc{Suh, E.~B.} (2002). A latent
autoregressive model for longitudinal binary data subject to informative
missingness. \textit{Biometrics} \textbf{58} 631--642.
\MR{1933536}

%b2 ###
\bibitem[\protect\citeauthoryear{}{1961}]{Bahadur1961}
\textsc{Bahadur, R.~R.} (1961). A representation of the joint
distribution of
responses to n dichotomous items. In \textit{Studies in Item Analysis and
Prediction} (H. Solomon, ed.).
\textit{Stanford Mathematical Studies
in the
Social Sciences VI} 158--168. Stanford Univ. Press.
\MR{0121893}

%b3 ###
\bibitem[\protect\citeauthoryear{}{2007}]{Caffo2007}
\textsc{Caffo, B.}, \textsc{An, M.-W.} and \textsc{Rohde, C.}
(2007). Flexible random intercept models for
binary outcomes using mixtures of normals. \textit{Comput.
Statist.
Data Anal.} \textbf{51} 5220--5235.
\MR{2370867}

%b4 ###
\bibitem[\protect\citeauthoryear{}{2006}]{Caffo2006}
\textsc{Caffo, B.} and \textsc{Griswold, M.} (2006). A user-friendly
introduction to
link-probit-normal models. \textit{Amer. Statist.} \textbf{60} 139--145.
\MR{2224211}

\bibitem[\protect\citeauthoryear{}{2002}]{D2002}
\textsc{Diggle, P. J., Heagerty, P., Liang, K. Y.} and \textsc{Zeger, S. L.}
(2002). \textit{Analysis of
Longitudinal Data}, 2nd ed.  Oxford Univ. Press,  Oxford.

%b5 ###
\bibitem[\protect\citeauthoryear{}{2001}]{Fahrmeir2001}
\textsc{Fahrmeir, L.} and \textsc{Tutz, G.} (2001). \textit
{Multivariate Statistical
Modelling Based
on Generalized Linear Models}. Springer, New York.
\MR{1832899}

%b6 ###
\bibitem[\protect\citeauthoryear{}{1995}]{Fitzmaurice1995}
\textsc{Fitzmaurice, G.~M.} (1995). A caveat concerning independence estimating
equations with multivariate binary data. \textit
{Biometrics} \textbf{51} 309--317.

\bibitem[\protect\citeauthoryear{}{1993}]{F1993}
\textsc{Fitzmaurice, G. M., Laird, N. M.} and \textsc{Rotnitzky, A. G.} (1993). Regression
models for discrete longitudinal responses (with discussion). \textit{Statist.
Sci.} \textbf{8}  248--309.

%b7 ###
\bibitem[\protect\citeauthoryear{}{1999}]{Heagerty1999}
\textsc{Heagerty, P.~J.} (1999). Marginally specified logistic-normal models
for longitudinal binary data. \textit{Biometrics} \textbf{55} 688--698.

%b8 ###
\bibitem[\protect\citeauthoryear{}{2000}]{Heagerty2000}
\textsc{Heagerty, P.~J.} and \textsc{Zeger, S.~L.} (2000).
Marginalized multilevel models
and likelihood inference (with comments and a rejoinder by the
authors). \textit{Statist. Sci.} \textbf{15} 1--26.
\MR{1842235}

%b9 ###
\bibitem[\protect\citeauthoryear{}{1971}]{Hoel1971}
\textsc{Hoel, P.~G.}, \textsc{Port, S.~C.} and \textsc{Stone, C.~J.}
(1971). \textit{Introduction
to Probability Theory}. Houghton Mifflin, Boston, MA.
\MR{0358880}

%b10 ###
\bibitem[\protect\citeauthoryear{}{2000}]{Hougaard2000}
\textsc{Hougaard, P.} (2000). \textit{Analysis of Multivariate
Survival Data}.
Springer, New York.
\MR{1777022}

%b11 ###
\bibitem[\protect\citeauthoryear{}{1997}]{Joe1997}
\textsc{Joe, H.} (1997). \textit{Multivariate Models and Dependence Concepts}.
Chapman and Hall, London.
\MR{1462613}

%b12 ###
\bibitem[\protect\citeauthoryear{}{1980}]{Kalbfleisch1980}
\textsc{Kalbfleisch, J.~D.} and \textsc{Prentice, R.~L.} (1980).
\textit{The Statistical Analysis of Failure Time Data}.
Wiley, New York.
\MR{0570114}

%b13 ###
\bibitem[\protect\citeauthoryear{}{1988}]{Laird1988}
\textsc{Laird, N.~M.} (1988). Missing data in longitudinal studies.
\textit{Stat. Med.} \textbf{7} 305--315.

%b14 ###
\bibitem[\protect\citeauthoryear{}{2004}]{Lee2004}
\textsc{Lee, Y.} and \textsc{Nelder, J.~A.} (2004). Conditional and
marginal models:
Another review. \textit{Statist. Sci.} \textbf{19}
219--228.
\MR{2140539}

%b15 ###
\bibitem[\protect\citeauthoryear{}{1986}]{Liang1986}
\textsc{Liang, K.~Y.} and \textsc{Zeger, S.~L.} (1986). Longitudinal
data analysis using generalized linear models. \textit{Biometrika}
\textbf{73} 13--22.
\MR{0836430}

%b16 ###
\bibitem[\protect\citeauthoryear{}{1998}]{Lipshultz1998}
\textsc{Lipshultz, S.~E.}, \textsc{Easley, K.~A.}, \textsc{Orav,
E.~J.}, \textsc{Kaplan, S.},
\textsc{Starc, T.~J.}, \textsc{Bricker, J.~T.},
\textsc{Lai, W.~W.},
\textsc{Moodie, D.~S.}, \textsc{McIntosh, K.}, \textsc{Schluchter,
M.~D.} and \textsc{Colan, S.~D.} (1998). Left ventricular
structure and function in children infected with human immunodeficiency virus:
The prospective P2C2 HIV Multicenter Study. Pediatric Pulmonary and Cardiac
Complications of Vertically Transmitted HIV Infection (P2C2 HIV) Study Group.
\textit{Circulation} \textbf{97} 1246--1256.

%b17 ###
\bibitem[\protect\citeauthoryear{}{2000}]{Lipshultz2000}
\textsc{Lipshultz, S.~E.}, \textsc{Easley, K.~A.}, \textsc{Orav,
E.~J.}, \textsc{Kaplan, S.}, \textsc{Starc, T.~J.}, \textsc{Bricker, J.~T.},
\textsc{Lai, W.~W.},
\textsc{Moodie, D.~S.}, \textsc{Sopko, G.} and \textsc{Colan, S.~D.} (2000).
Cardiac dysfunction and mortality
in HIV-infected children: The Prospective P2C2 HIV Multicenter Study. Pediatric
Pulmonary and Cardiac Complications of Vertically Transmitted HIV Infection
(P2C2 HIV) Study Group. \textit{Circulation} \textbf{102} 1542--1548.

%b18 ###
\bibitem[\protect\citeauthoryear{}{2002}]{Lipshultz2002}
\textsc{Lipshultz, S.~E.}, \textsc{Easley, K.~A.}, \textsc{Orav,
E.~J.}, \textsc{Kaplan, S.}, \textsc{Starc, T.~J.}, \textsc{Bricker, J.~T.},
\textsc{Lai, W.~W.},
\textsc{Moodie, D.~S.}, \textsc{Sopko, G.}, \textsc{Schluchter,
M.~D.} and \textsc{Colan, S.~D.} (2002). Cardiovascular
status of
infants and children of women infected with HIV-1 (P(2)C(2) HIV): A~cohort
study. \textit{Lancet} \textbf{360} 368--373.

%b19 ###
\bibitem[\protect\citeauthoryear{}{1991}]{Lipsitz1991}
\textsc{Lipsitz, S.~R.}, \textsc{Laird, N.~M.} and \textsc
{Harrington, D.~P.}
(1991). Generalized estimating equations for correlated binary
data: Using the odds ratio as a measure of association. \textit
{Biometrika} \textbf{78} 153--160.
\MR{1118240}

%b20 ###
\bibitem[\protect\citeauthoryear{}{1989}]{McCullagh1989}
\textsc{McCullagh, P.} and \textsc{Nelder, J.~A.} (1989). \textit{Generalized
Linear Models}, 2nd ed. Chapman and Hall, New York.
\MR{0727836}

%b21 ###
\bibitem[\protect\citeauthoryear{}{1994}]{Molenberghs1994}
\textsc{Molenberghs, G.} and \textsc{Lesaffre, E.} (1994). Marginal
modelling of
correlated ordinal data using a multivariate Plackett distribution.
\textit{J. Amer. Statist. Assoc.} \textbf{89} 633--644.

%b22 ###
\bibitem[\protect\citeauthoryear{}{1999}]{Nelson1999}
\textsc{Nelsen, R.~B.} (1999). \textit{An Introduction to Copulas.}
Springer, New York.
\MR{1653203}

%b23 ###
\bibitem[\protect\citeauthoryear{}{1991}]{Neuhaus1991}
\textsc{Neuhaus, J.~M.}, \textsc{Kalbfleisch, J.~D.} and \textsc
{Hauck, W.~W.} (1991). A comparison
of cluster-specific and
population-averaged approaches for analyzing correlated binary data.
\textit{Int. Statist. Rev.} \textbf{59} 25--35.

%b24 ###
\bibitem[\protect\citeauthoryear{}{1995}]{Pinheiro1995}
\textsc{Pinheiro, J.~C.} and \textsc{Bates, D.~M.} (1995).
Approximations to the
log-likelihood function in the nonlinear mixed-effects model. \textit
{J.~Comput. Graph. Statist.} \textbf{4} 12--35.

%b25 ###
\bibitem[\protect\citeauthoryear{}{1976}]{Rubin1976}
\textsc{Rubin, D.~B.} (1976). Inference and missing data.
\textit{Biometrika} \textbf{63} 581--592.
\MR{0455196}

%%b26 ###
%J.~H.} (1984). Random-effects
%models for serial observations with binary response. \textit
%{Biometrics} \textbf{40}
%961--971.

%b27 ###
\bibitem[\protect\citeauthoryear{}{2003}]{Wang2003}
\textsc{Wang, Z.} and \textsc{Louis, T.~A.} (2003). Matching
conditional and marginal
shapes in binary mixed-effects models using a bridge distribution
function. \textit{Biometrika} \textbf{90} 765--775.
\MR{2024756}

%b28 ###
\bibitem[\protect\citeauthoryear{}{2004}]{Wang2004}
\textsc{Wang, Z.} and \textsc{Louis, T.~A.} (2004). Marginalized
binary mixed-effects
with covariate-dependent random effects and likelihood inference.
\textit{Biometrics} \textbf{60} 884--891.
\MR{2133540}

%b29 ###
\bibitem[\protect\citeauthoryear{}{1990}]{Zhao1990}
\textsc{Zhao, L.~P.} and \textsc{Prentice, R.~L.} (1990). Correlated
binary regression using a quadratic exponential model. \textit
{Biometrika} \textbf{77} 642--648.
\MR{1087856}
\end{thebibliography}
\end{document}